\documentclass[preprint,12pt]{elsarticle}

\usepackage{amsmath,amssymb,amsfonts}
\usepackage{multicol,caption}
\usepackage{commath}
\usepackage{algorithm,algorithmic}
\usepackage{graphicx}
\usepackage{caption}
\usepackage{subcaption}
\usepackage{booktabs,ragged2e}
\usepackage{tabularx}
\usepackage{textcomp}
\usepackage{xcolor}
\usepackage{hyperref}
\usepackage{fancyhdr}
\usepackage{tikz-uml}
\tikzumlset{font=\scriptsize\sffamily}
\usepackage{pgfplots}
\pgfplotsset{compat=1.8}
\usepgfplotslibrary{statistics}
\usepackage[T1]{fontenc}
\usepgfplotslibrary{colorbrewer}
\pgfplotsset{compat = 1.15, 
             cycle list/Dark2} 
\def\BibTeX{{\rm B\kern-.05em{\sc i\kern-.025em b}\kern-.08em
    T\kern-.1667em\lower.7ex\hbox{E}\kern-.125emX}}

\begin{document}

\begin{frontmatter}



\title{PeerFL: A Simulator for Peer-to-Peer Federated Learning at Scale}


\author[inst1]{Alka Luqman}

\affiliation[inst1]{organization={Nanyang Technological University},
            addressline={Singapore}}

\author[inst2]{Shivanshu Shekhar}
\author[inst1]{Anupam Chattopadhyay}

\affiliation[inst2]{organization={Indian Institute of Technology},
            addressline={Madras}, 
            country={India}}

\begin{abstract}
This work integrates peer-to-peer federated learning tools with NS3, a widely used network simulator, to create a novel simulator designed to allow heterogeneous device experiments in federated learning. This cross-platform adaptability addresses a critical gap in existing simulation tools, enhancing the overall utility and user experience. NS3 is leveraged to simulate WiFi dynamics to facilitate federated learning experiments with participants that move around physically during training, leading to dynamic network characteristics. Our experiments showcase the simulator's efficiency in computational resource utilization at scale, with a maximum of 450 heterogeneous devices modelled as participants in federated learning. This positions it as a valuable tool for simulation-based investigations in peer-to-peer federated learning. The framework is open source and available for use and extension to the community. 
\end{abstract}




\end{frontmatter}


\section{Introduction}
\label{sec:introduction}

\subsection{Motivation}
The convergence of machine learning and decentralized edge networks is ushering in a new era of technology, where intelligent agents can collaborate, learn and adapt in a distributed and scalable manner. Federated learning (FL) \cite{1902.04885} offers the potential to harness the collective intelligence of connected devices and systems while preserving individual privacy \cite{truong2021privacy}, thus enabling applications in diverse fields such as IoT\cite{ASGHARI2019241, 2104.07914}, autonomous systems\cite{zeng2022federated}, and healthcare diagnostics\cite{roy2019braintorrent}. Peer-to-peer federated learning (P2P FL) is a decentralized version of FL where learning must occur over multiple non-trusted devices. The effective deployment and optimization of P2P FL algorithms in real-world scenarios present significant challenges due to the complex and dynamic nature of the network in which it is deployed and its communication environments \cite{kairouz2021advances}. \par
To address these challenges, we introduce a novel simulator that integrates FL training with network simulation, offering a robust platform for the exploration of P2P FL algorithms in various network and communication settings. The intent was to model and explore the impact to peer-to-peer federated learning when the participants are mobile and move around with respect to their WiFi base stations. This dynamism affects communication latencies, and thereby time taken to converge to an acceptably good global or personal model. Our work is motivated by the growing demand for understanding how P2P FL models perform under diverse conditions and the need for a unified framework to assess their adaptability, scalability, and robustness in decentralized environments. \par

\subsection{Contribution}
The primary contribution of this work is to present a comprehensive simulator that seamlessly integrates P2P FL algorithms with network and communication simulation, enabling researchers and practitioners to assess the behaviour of P2P FL algorithms in various network and learning settings. This framework offers the flexibility to model various network topologies and dynamic network conditions and to easily scale up experiments, thereby serving as a valuable tool for exploring the interaction between federated learning objectives and network dynamics in resource-constrained settings. \par

The key advantages to this simulator are:
\begin{itemize}
    \item \textbf{Realistic Network Dynamics}: PeerFL creates realistic network environments by incorporating factors such as latency, bandwidth, and packet loss while modelling the network topology. This allows for the evaluation of P2P FL algorithms in conditions that mirror real-world scenarios like sudden bandwidth drops that increase communication costs, devices dropping out of training rounds and its impact on the overall learning objective etc. This is important to create adaptable learning algorithms under different network constraints.\par
    

    \item \textbf{Isolated and Scalable Testing}: ocker's containerization capabilities provides the user with isolated environments for running P2P FL experiments. Researchers can deploy multiple containers representing different FL participants with different hardware capabilities. This approach allows for scalable testing and the evaluation of algorithms under varying device conditions and network topologies. It is also possible to conduct all experiments with and without network simulation, to help validate the complexities introduced by network dynamics.


    \item \textbf{Open-Source Framework}: We offer the simulator as an open-source framework, fostering collaboration within the research community and providing a foundation for further innovation and development in the field of P2P FL in real-world resource-constrained networks.
    
\end{itemize}

\subsection{Organization of the Paper}
The paper introduces the background of federated learning and peer-to-peer federated learning along with related works and existing frameworks to simulate federated learning in Section \ref{sec:literature}. The system design, different usage models and the simulator architecture is presented in Section \ref{sec:system-design}. Further implementation details and workflows are given in Section \ref{sec:implementation} along with examples for reference. The framework evaluation benchmarks are presented in Section \ref{sec:evaluation} before concluding and discussing the simulator's future roadmap in Section \ref{sec:conclusion}.

\section{Background and Related Works}
\label{sec:literature}

\subsection{Federated Learning}
Data is the best regulariser for any machine learning model. Machine learning models have shown great success with the use of big data, and many people hope to use big data to boost the quality of AI models. However, this is rarely realized in practice, as most industries are restricted by legal regulations \cite{doi:10.1080/13600834.2019.1573501, CHIK2013554, privacy_gaff} and are bound to process their data in silos. \par
\begin{figure}[H]
        \begin{subfigure}[t]{0.35\textwidth}
        \centering
        \includegraphics[scale=0.4]{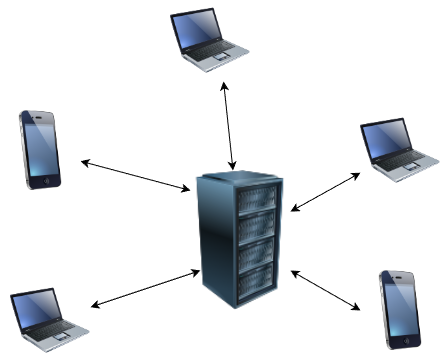}
        \caption{Client-Server}
        \label{fig:cs_fl}
        \end{subfigure}
    \hspace{2em}
    \begin{subfigure}[t]{0.35\textwidth}
    \centering
    \includegraphics[scale=0.4]{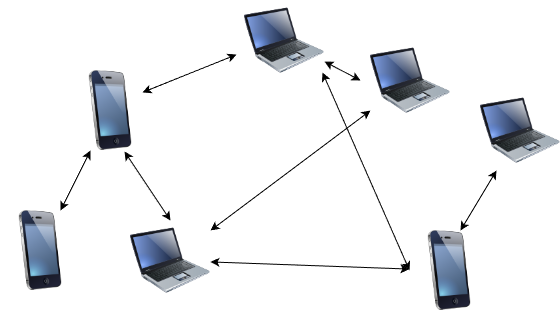}
    \caption{Peer-to-Peer}
    \label{fig:p2p_fl}
    \end{subfigure}
    \caption{An illustration of the main variants of federated learning.}
    \end{figure}

The concept of federated learning \cite{mcmahan2017communicationefficient} was first proposed to develop machine learning models using data sets distributed across multiple devices while safeguarding against data leakage. There are two main variants of federated learning seen nowadays.
\begin{itemize}
    \item Client-Server FL: Every client that participates in the training will train AI models on its own private data and send the model weights after encryption to a central server, as shown in Figure \ref{fig:cs_fl}. This trusted server is responsible for aggregating all the individual models to make a global model, which is then transmitted back to the clients for use or further rounds of training.
    \item Peer-to-Peer FL: A fully decentralized variant is depicted in Figure \ref{fig:p2p_fl} where there is no trusted aggregator. This requires learning to occur using peer-to-peer communication between individual clients. Each participant device performs local training as well as transmits their model to their peers. Every peer that receives another model can perform aggregation, fine-tuning and further transmission of models.
\end{itemize}
    
In this paper, we focus on the peer-to-peer federated learning setting. The network connectivity between devices, along with the sparsity of the underlying graph influences the learning process.  It is also important to note that there is no longer a global state of the model as in standard federated learning since devices can always choose to continue fine-tuning or personalization of their models. \par

\subsection{Peer-to-Peer Federated Learning}
Peer-to-peer federated learning (P2P FL) is a developing field of study that intends to distribute the federated learning training process across participating peers, enabling cooperative learning without the need for a centralized server. While this is a relatively new field, there are some notable works and approaches that have contributed to its development. We present a few prominent examples below.
\begin{itemize}
    \item \textbf{Peer-averaging(PA) algorithm} \cite{mcmahan2017communicationefficient}: The peer-averaging algorithm, proposed by McMahan et al. in their paper "Communication-Efficient Learning of Deep Networks from Decentralized Data," introduced the concept of decentralized federated learning. In this approach, participating peers (devices) collaborate to train a shared model by exchanging model updates and averaging them locally. This method reduces the reliance on a central server and allows training to occur directly between peers.
    
    \item \textbf{FedP2P} \cite{2203.12285}: FedP2P, introduced by Zhao et al. in their paper "Federated Peer-to-Peer Learning," focuses on a fully decentralized P2P FL framework. It employs a gossip-based communication protocol, where peers exchange model updates with a subset of other peers. This approach aims to enhance scalability and robustness in large-scale P2P FL systems by mitigating the issues associated with centralized coordination and communication bottlenecks.

    \item \textbf{Secure P2P FL} \cite{2101.12316}: Several works have focused on ensuring privacy and security in P2P FL settings. For instance, the work "Privacy-Preserving Decentralized Machine Learning with Byzantine Robust Federated Averaging" by Gupta et al. proposes a Byzantine-robust P2P FL algorithm that can withstand malicious peers. They employ cryptographic techniques, such as secure multiparty computation (MPC), to enable secure aggregation of model updates without compromising privacy.
\end{itemize}

\par
Peer-to-Peer Federated Learning (P2P FL) is different from Distributed Machine Learning (DML) in its communication pattern and coordination. In P2P FL, each participating device has more autonomy and control over the training process. Devices collaborate directly with each other, making decisions about which peers to communicate with and when to exchange model updates. In DML, the centralized coordinator or parameter server plays a crucial role in coordinating the training process, distributing tasks, and aggregating model updates. This is responsible for the high scalability and fault-tolerant potential of P2P FL. \par
Since P2P FL avoids the reliance on a central server for coordination, it can distribute the computational and communication load across participating devices, potentially enabling more efficient and scalable training. If a peer in the P2P network fails or becomes unavailable, other peers can continue the training process by collaborating with different peers. This decentralized nature helps in building robust systems that can handle node failures.\par

\subsection{Frameworks in FL}


From a systems perspective, the lack of frameworks that facilitate scalable FL training on mobile and edge devices is a significant obstacle to P2P FL research. Towards this end, the \textbf{Flower framework} \cite{2007.14390} was developed for simulating the Client-Server setting of federated learning. It offered a stable, language and ML framework-agnostic implementation of the core components of an FL system hence offering higher-level abstractions so that researchers may experiment with and put new concepts into practice on top of a solid stack. Flower provides a distributed infrastructure for FL, allowing the training of ML models across a decentralized network of devices or servers. It supports different communication protocols, including web sockets and gRPC, to facilitate the communication between clients and the server. Flower follows a client-server architecture \cite{2104.03042} and offers flexibility in defining aggregation strategies for combining model updates from multiple clients. \par

\textbf{FATE} \cite{liu2021fate} is another framework that allows users to configure different FL algorithms along with security implementations of homomorphic encryption and multi-party computation. It is geared towards industries which have multiple organizations that can collaborate to learn from distributed data. \par

\textbf{FLSim} \cite{githubGitHubFacebookresearchFLSim} is a library that supports various FL aggregation algorithms in a centralized FL setting for collaborative training. It incorporates some compression techniques to allow ML training on resource-constrained devices like mobile phones. \par

Although many frameworks are emerging for federated learning, the same can't be said about the peer-to-peer decentralized federated learning domain. Towards this end, we are proposing PeerFL which apart from being able to support various FL algorithms in the P2P setting, also gives the freedom to change the network and channel properties via the NS3 network simulator. We leverage many of the common transmission models already implemented by the NS3 framework to provide this feature in PeerFL.\par

\subsection{Network Simulator : NS3}
NS3 is a discrete-event network simulator for Internet systems, targeted primarily for research and educational use \cite{article}. A full-featured TCP/IP protocol stack is available, and various wireless technologies, including LTE and WiFi, are supported through the general-purpose network simulator NS3. \par

Key features and characteristics of NS3 include:
\begin{itemize}
    \item \textbf{Protocol Modeling}: NS3 provides a comprehensive set of built-in models for various network protocols, such as TCP/IP, UDP, ICMP, routing protocols (e.g., OSPF, BGP), wireless protocols (e.g., IEEE 802.11, LTE), and more. Users can also create custom protocol models to simulate specific scenarios or evaluate new protocols.
    
    \item \textbf{Flexible Scenario Creation}: NS3 offers flexibility in creating network scenarios by allowing users to define network topologies, traffic patterns, and application behaviors. It supports the simulation of diverse network environments, mobility scenarios, and network failures.
    
    \item \textbf{Realistic Network Simulation}: NS3 allows users to create realistic network simulations by modeling network nodes, links, routers, and devices. It supports a wide range of network technologies, including wired and wireless networks, Internet protocols, and communication standards.
\end{itemize}
NS3 supports TAP/TUN devices to support integration with testbeds and real applications. The TAP (network tap) device is a virtual network interface that allows communication between the simulated network in NS3 and the host operating system or other external networks. It acts as a bridge between the NS3 simulation and the real network stack. However, it is to be noted that NS3 loss models can be very extensive on the compute resources \cite{inbook} and was optimized to a certain degree for use in PeerFL.

\section{System Design}
\label{sec:system-design}
\subsection{Design Principles}
To ensure effective scalability and easy usage of the simulator, the following design principles were of prime importance while creating PeerFL:
\begin{itemize}
    \item \textbf{Modularity}: PeerFL should be modular so that experts in different domains can easily add or modify components, such as learning algorithms, network models, and communication protocols. This promotes flexibility and encourages the integration of new technologies and algorithms as they emerge, without requiring extensive engineering efforts.
    \item \textbf{Scalability and Parallelization}: To enable realistic research on a large scale, PeerFL should be capable of handling a significantly large number of clients \textit{concurrently}, as real-world applications would involve learning from numerous clients. To handle such large-scale simulations efficiently, PeerFL should support parallel execution and leverage optimization techniques to maximize performance. This is particularly important for simulating complex P2P FL scenarios.
    \item \textbf{Realism}: PeerFL should possess the flexibility to accommodate the dynamic nature of clients in P2P FL and the rapidly evolving ML ecosystem. The closer the simulator mirrors real-world conditions, the more valuable it becomes for research and testing. This includes support for dynamic experiment settings which modify the network parameters over the life cycle of a single test run. Another aspect to maintaining fidelity with the real world is to allow the capability of modelling adversarial clients in addition to the honest devices participating in P2P FL. Since FL algorithms are primarily used in situations that involve sensitive data or have privacy considerations, the simulator should include provisions for modelling attacks and implementing security defences within the simulations.
    \item \textbf{Client-agnostic Implementation}: In order to accommodate the diverse range of mobile client environments, PeerFL should be designed to be compatible with diverse hardware. This interoperability is crucial since different devices or agents seldom have the same hardware in the real world. This further allows exploration of the impact of device heterogeneity, which is an important area of study in P2P FL.
\end{itemize}

\begin{figure*}[!h]
\begin{center}
\begin{tikzpicture}
\catcode`\_=12 

\begin{umlpackage}[x=0,y=1]{1}
\umlclass{Channel}{
  data_rate : double \\ delay : int \\ packet : numpy.ndarray 
}{
  \umlvirt{setDataRate() : None} \\ 
  \umlvirt{setDelay() : None} \\
  \umlvirt{setPacketContent(Tensor) : Bool}
  }
\end{umlpackage}
\begin{umlpackage}[x=5,y=1]{2}
\umlinterface{Device}{
  local_data : uint \\ device_attributes : string \\
  privacy_attributes : DP \\ topology : p2p
}{
\umlvirt{CreateData(n) : Bool} \\
\umlvirt{LoadData() : Bool} \\
\umlvirt{PrepData() : Bool} \\
\umlvirt{BatchData() : Bool} \\
}
\end{umlpackage}
\begin{umlpackage}[x=10,y=1]{3}
\umlclass{Model}{
  keras_model : tf \\ epochs : int \\
  optimizer : tf.keras \\ learning_rate : double \\ metrics : ndarray
}{
\umlvirt{InitializeModel() : keras_model} \\
\umlvirt{Train() : metrics} \\
\umlvirt{Evaluate() : metrics} 
}
\end{umlpackage}

\begin{umlpackage}[x=0,y=0]{4}
\umlclass[x=4,y=-5]{Client}{
  device : Device\\ local_model : Model \\ comm : Channel
}{
\umlvirt{UpdateModel() : Bool} \\
\umlvirt{Communicate() : Bool} 
}

\umlclass[x=2,y=-10]{WiFiClient}{
  comm : Channel \\ channel_helper : ns3.YansWifiChannelHelper \\
  dev : ns3.NetDeviceContainer
}{
\umlvirt{CreateCommsContent() : ns3.packet} \\
\umlvirt{Communicate() : Bool} 
}
\umlclass[x=10,y=-10]{ZMTPClient}{
  comm : Channel \\ outgoing_socket : SocketType.DEALER \\
  incoming_socket : SocketType.ROUTER
}{
\umlvirt{CreateCommsContent() : dict} \\
\umlvirt{Communicate() : Bool} 
}
\end{umlpackage}

\umlinherit[geometry=-|]{Client}{1}
\umlinherit[geometry=-|]{Client}{2}
\umlinherit[geometry=-|]{Client}{3}

\umlassoc[geometry=-|-, mult1=1, pos1=0.3, arg2=0..*, pos2=1.55]{Client}{ZMTPClient}
\umlassoc[geometry=-|-, mult1=1, pos1=1.45, arg2=0..*, pos2=1.56]{Client}{WiFiClient}

\end{tikzpicture}
\end{center}
\caption{Class diagram of PeerFL describing the structure and behaviour of the system.}
\label{fig:class_diagram}
\end{figure*}
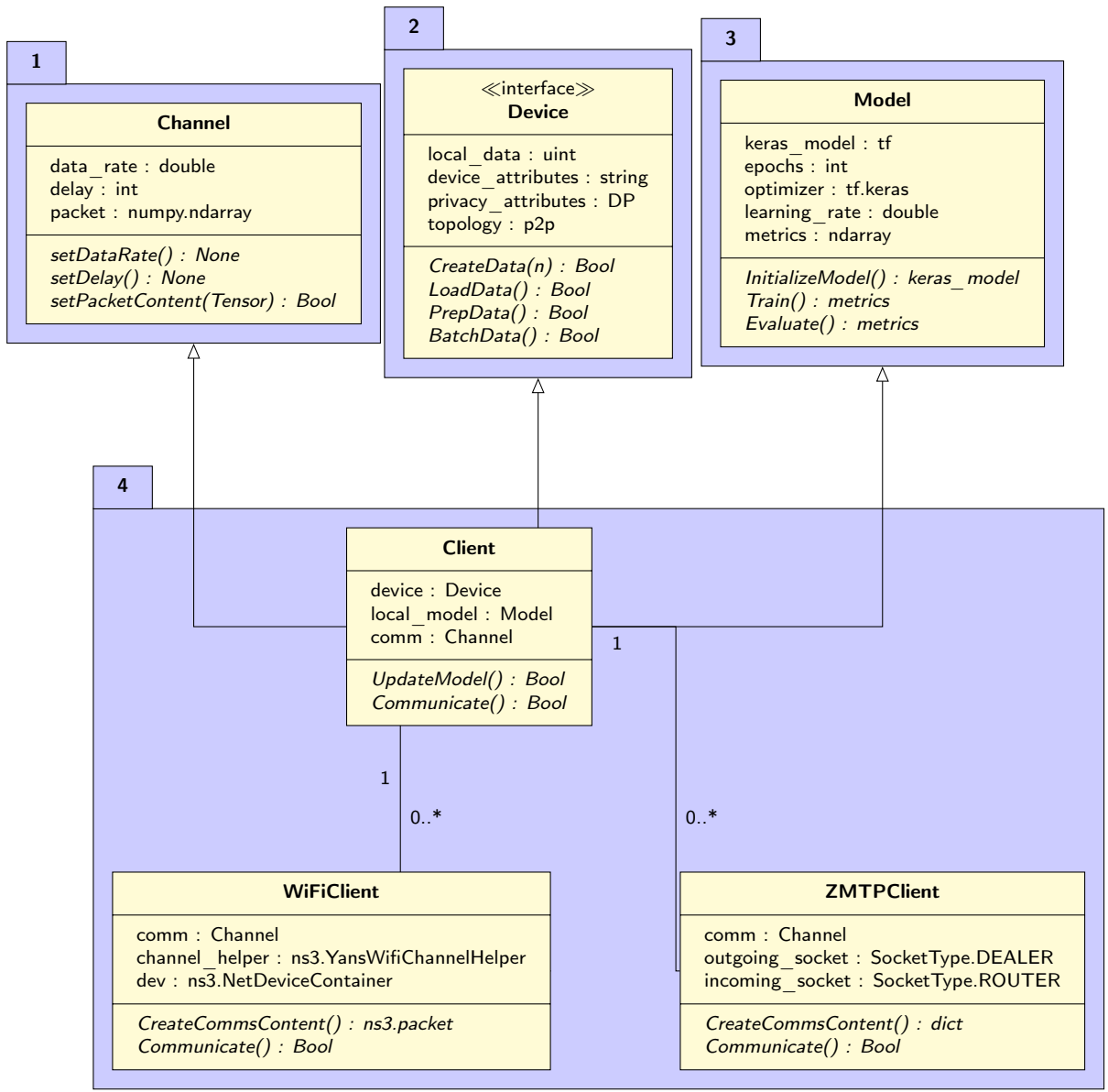

An architecture that accommodates these properties will enhance the realism and scalability of P2P FL research, facilitating a seamless transition from simulated experiments to conducting large-scale research on actual edge devices. The class diagram of PeerFL is shown in Figure \ref{fig:class_diagram} to highlight how abstraction and inheritance are used to implement these properties. \par

\subsection{Usage Models}
Three main categories of experiments can be performed on the simulator.
\begin{itemize}
    \item \textbf{Experiments at scale:} To gain a deeper understanding of the generalizability of methods in P2P FL, experiments should be able to scale effectively with both a large pool of clients and a substantial number of clients training concurrently. PeerFL was tested on upto 450 concurrent devices to ensure that it is capable of conducting large-scale evaluations of algorithms and designs, utilizing reasonable compute resources such as a single machine or a multi-GPU rack. Additionally, the results obtained at this scale are produced within an acceptable time frame, ensuring efficient execution in terms of wall-clock execution time.
    \item \textbf{Experiments on heterogeneous hardware:} P2P FL research commonly encounters heterogeneous client environments. To effectively study the impact of system heterogeneity, researchers can specify different operating systems for their devices, which are implemented using different Docker base images. They can also specify RAM or bandwidth limitations and GPU non-availability to further constrain the experimentation environment. PeerFL deploys heterogeneous experiments in a seamless process and makes it straightforward to collect performance measurements across all these clients.
    \item \textbf{Experiments on Security:} By incorporating both honest and malicious nodes in PeerFL, researchers can study the impact of various attack scenarios, evaluate the resilience of the model against adversarial behaviour, and develop robust mechanisms for detecting and mitigating attacks. Understanding the behaviour and interactions of both honest and malicious nodes is crucial for advancing the security and effectiveness of P2P FL algorithms.
\end{itemize}

\subsection{System Architecture}

\begin{figure*}[!t]
    \centering
    \includegraphics[width=\textwidth]{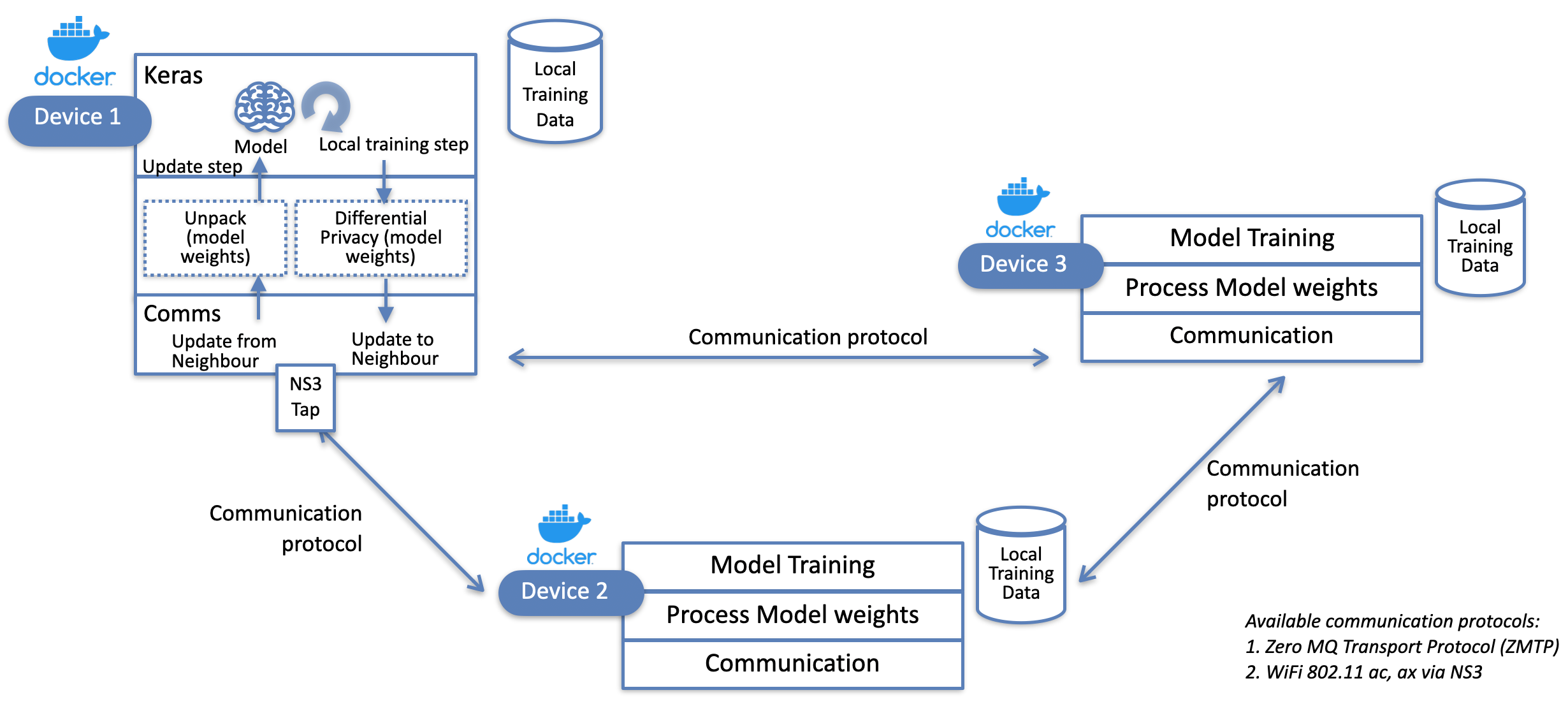}
    \caption{PeerFL Architecture diagram}
    \label{fig:architecture}
\end{figure*}

For the implementation of PeerFL, we opted to use Python as the programming language along with TensorFlow for the model building. Python provided us with the necessary flexibility in asynchronous processing which is required to perform concurrent training and peer communication in the simulator. Lightweight containerization using Docker was used to model individual devices. The containers were configured via YAML to restrict the availability of RAM, bandwidth and hardware acceleration for specific devices. Kubernetes was used to create subnets and manage the dockers. This allowed 2 ways of communication between the containers - either using ZeroMQ (ZMTP) without mobility of devices or a simulated network facilitated by NS3 with more control over the devices in the communication network. These communication channels allowed the devices to exchange data and information required for the P2P FL training process.\par

Figure \ref{fig:architecture} shows a high-level view of PeerFL. Each device can be envisioned as having three layers in its functionality stack. The lowest layer is in charge of communication-related functions - including sending and receiving messages/packets to its peers. It also applies commonly used compression techniques to save network bandwidth usage. The middle layer unpacks the received messages into compliant data structures which can be easily used for ML model training. This layer abstracts a lot of the complexity to allow the inter-processing of tensors and network packets. The upper layer focuses on training the model using a chosen FL algorithm. Any FL algorithm can be tested on the simulator with minimal changes to perform the distributed aggregation across all peers. \par

TAP (network tap) and TUN (network tunnel) devices are virtual network interfaces commonly used in networking and software-defined networking (SDN) environments.\par
\textbf{TAP Devices} simulate a network tap, which is a device used for monitoring or capturing network traffic. TAP devices are often employed in network analysis, packet sniffing, or traffic monitoring applications. When a network interface is set as a TAP device, it operates in promiscuous mode, allowing it to capture and monitor all network traffic passing through the interface. TAP devices provide a copy of the network traffic to applications or tools for analysis, without interrupting the flow of traffic to its intended destination. \par
\textbf{TUN Devices}, also known as network tunnelling devices, are virtual network interfaces used for creating network tunnels or encapsulating network packets within another protocol. TUN devices allow applications or operating systems to establish virtual private network (VPN) connections, implement network overlays, or create secure communication channels. Network packets sent to a TUN device are encapsulated within another protocol (such as IP over IP or IP over UDP) and then forwarded to the appropriate destination.\par

Both TAP and TUN devices are widely used in various networking scenarios, including virtualization, VPNs, SDNs, etc. Currently, NS3 provides support for CSMA (Carrier Sense Multiple Access) and WiFi (Wireless Fidelity) networks in the context of TAP simulation. NS3 allowed us to emulate dynamic WiFi behaviour like the mobility of participant devices. This means that we were able to simulate the characteristics and performance of these networks and then perform the P2P FL aggregation mechanisms. We could control factors such as signal propagation, interference, and packet losses, enabling us to investigate the robustness and resilience of P2P FL algorithms in different network environments.\par
By incorporating the NS3-simulated network into our framework, we gained significant flexibility in terms of simulating various network scenarios and customizing channel losses. Furthermore, as NS3 continues to evolve and expand its capabilities, we have the opportunity to incorporate additional types of networks into our framework. Whenever NS3 introduces support for new network types, we can integrate them into PeerFL, further enhancing the realism and versatility of P2P FL research.\par

\section{Implementation}
\label{sec:implementation}
Each Docker container in PeerFL operates independently with its own operating system. Each container has its own network stack, allowing it to function as a separate entity. Communication between containers is facilitated through a network device (eth0), which acts as the interface for transmitting and receiving network traffic. To establish connectivity with the host's operating system, Linux bridges are utilized. These bridges serve as connections between the individual container's eth0 devices and the host operating system. This arrangement enables seamless communication and data transfer between multiple containers on a single host and across multiple hosts. This is what allows the scalability of experiments in a distributed manner without requiring huge powerful machines.\par
To capture and analyze network packets, TAP devices are employed. These TAP devices are connected to the Linux bridges, acting as interceptors for network traffic passing through them. As packets flow through these TAP devices, they show up in the overarching Kubernetes network, where they can be intercepted and analyzed using NS3. A specialized Net device in NS3 is responsible for connecting to the TAP devices and forwarding the captured packets to an NS3 ``virtual node". This allows NS3 to restrict the network characteristics, enabling us to model, observe and study the flow of packets within the dynamically simulated environment. \par

The NS3 virtual node serves as a proxy, representing the Docker container within the NS3 simulation environment. It plays a crucial role in facilitating communication between the container and the simulated network. When packets are received through the network TAP, they are forwarded to the corresponding WiFi Net device within NS3. Similarly, packets that enter the NS3 simulation through the WiFi net device are forwarded back to the network TAP, providing the illusion to the Docker container and its application that it is connected to a WiFi network. To enable communication between containers and NS3, IP addresses are created dynamically at start up and assigned to each container. By assigning these IP addresses, NS3 is aware of the simulated node addresses, enabling the use of TCP/IP protocols within the simulation. This IP assignment allows us to specify peers that each device is directly connected to, for sending and receiving data directly. If a device is not a direct peer, the path to the required peer is found from a global adjacency matrix and traversed to reach the destination. While NS3 provides the foundational code for simulating the WiFi network, some modifications were made to support TCP/IP protocols. These changes involved updating the code to assign the simulated nodes their respective addresses within NS3. These modifications allow for the effective simulation of TCP/IP-based communication, enabling the use of protocols like TCP and IP within the FL framework. Training rounds were decoupled from the communication, as shown in the architecture (Figure \ref{fig:architecture}). These operations were scheduled in an asynchronous manner to achieve parallel computation and communication, eliminating unnecessary waiting time.\par

PeerFL supports both centralized and P2P settings of FL. For the centralized setting the user can specify the node that they want to act as a trusted aggregator and when they run the simulator the terminal get attached to the server, whereas in P2P the user can specify the entire training graph and the terminal will get attached to the last node receiving the model weights.\par

We summarize the entire flow for a sample PeerFL simulation for an image classification task. The simulation is initialized (Algorithm \ref{algo:driver}) and begins with one node training a classification model on its own local data. The trained neural network's weights are extracted, compressed and packaged according to the communication protocol chosen. This node then transmits the message to its next peer(s), as detailed in Algorithm \ref{algo:indevice}. This process is repeated until the trained model's accuracy saturates. This is identified by monitoring model loss continuously and using known early stopping techniques \cite{prechelt2002early}. \par

\begin{algorithm}
\caption{Driver Code}
\begin{algorithmic}[1]
  \REQUIRE YAML Configuration file
  \STATE Initialize the Docker containers using specified configuration
  \STATE Initialize the Kubernetes network to assign IP addresses to each device
  \STATE Initialize the Ns3 network using TAP and above assigned IP addresses
  \STATE Execute Training Code (Algorithm \ref{algo:indevice}) in each device
  \STATE Append received ML Accuracy to Early Stopping Daemon
\end{algorithmic}
\label{algo:driver}
\end{algorithm}

\begin{algorithm}
\caption{Training Code}
\begin{algorithmic}[1]
    \IF{Node Exists}
      \STATE Do Nothing
    \ELSE
      \STATE Initialize the local Node
    \ENDIF
  \FOR{each iteration}
    \IF{receive}
      \STATE Poll until connection is established
      \STATE Receive and unpack neural network weights
      \STATE Average local neural network weights with the received neural network weights
      \STATE Train new model on local data 
      \STATE Extract neural network weights 
      \STATE Send weights to peer node
    \ELSE
      \STATE Initialize local model with random weights 
      \STATE Train local model on local data
      \STATE Extract neural network weights 
      \STATE Send weights to peer node
    \ENDIF
  \ENDFOR
  \RETURN Time Taken for above procedure, ML Accuracy
\end{algorithmic}
\label{algo:indevice}
\end{algorithm}

\begin{figure*}[!]
    \centering
    \includegraphics[width=0.8\textwidth]{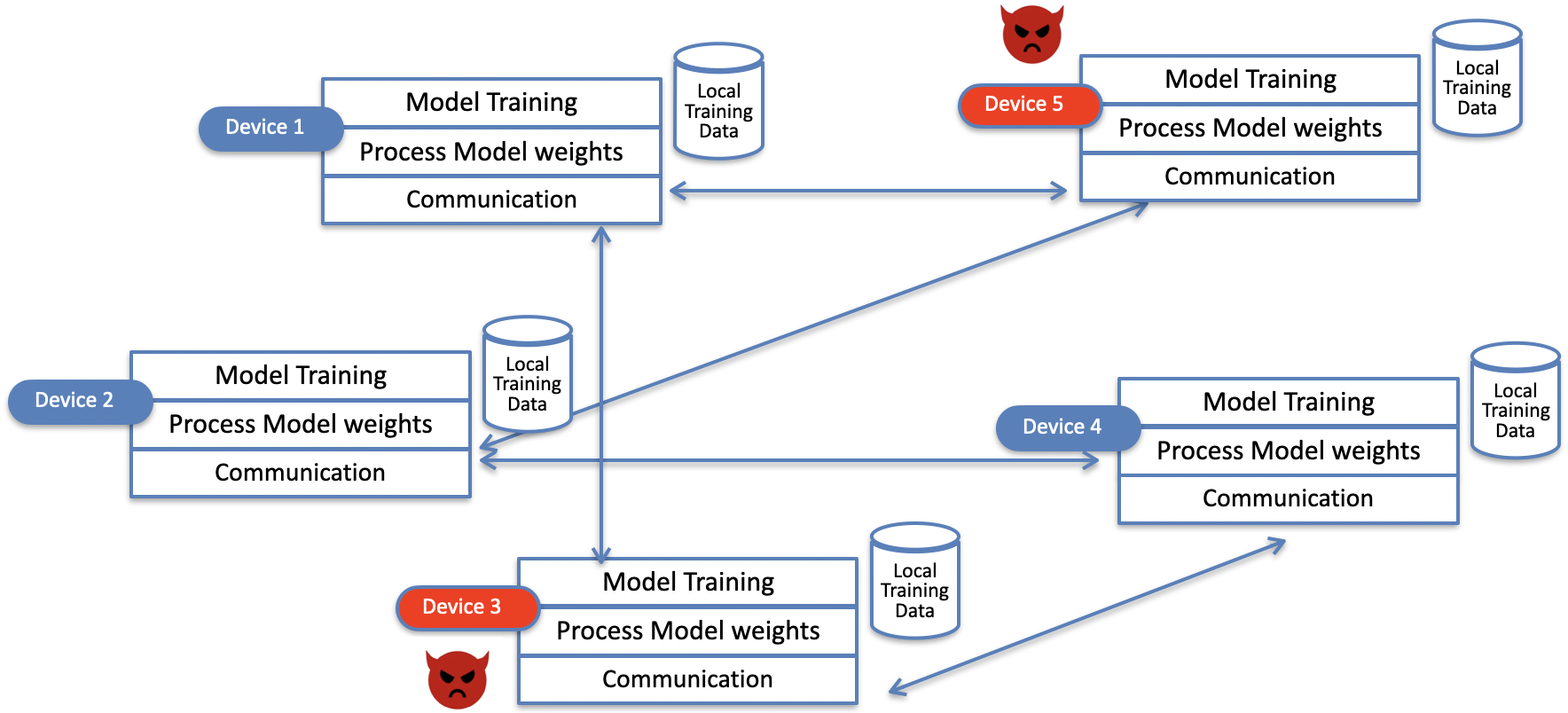}
    \label{fig:attack_flsim}
    \caption{Modelling attacks in P2P FL using PeerFL}
\end{figure*}

In NS3, the TAP (network tap) device is often used to bridge the simulated network with the host operating system or external networks. However, it is worth noting that using TAP devices in NS3 simulations can sometimes introduce performance overhead, leading to slower simulation execution. The slowness of TAP devices in NS3 simulations can be attributed to several factors:
\begin{itemize}
    \item \textbf{Packet Processing Overhead}: TAP devices require packet processing between the simulated network in NS3 and the host operating system. Each packet that enters or leaves the simulation needs to be processed by the TAP device, adding overhead to the simulation execution time. This packet processing can become a significant bottleneck, especially when dealing with a large number of packets or high network traffic.

    \item \textbf{Context Switching}: When using TAP devices, NS3 needs to communicate with the host operating system to send and receive network packets. This involves frequent context switching between the simulation and the operating system, which can introduce additional overhead and slow down the simulation.

    \item \textbf{System Configuration}: The performance of TAP devices in NS3 simulations can also be affected by the configuration of the host operating system and the specific network setup. Factors such as network buffer sizes, CPU scheduling, and system load can impact the efficiency of packet handling and overall simulation performance.
\end{itemize}

It is important to note that the performance of TAP devices in NS3 simulations can also depend on the specific hardware and software environment in which the simulations are run. Therefore, it is recommended to carefully profile and optimize the simulation setup based on the specific requirements and constraints of the simulation scenario.

\subsection{Modelling Attacks}

Incorporating both adversarial and honest-but-curious devices in PeerFL creates a dynamic and realistic testing environment for P2P FL algorithms. This approach enhances the utility of PeerFL for security-focused research. It enables the development of more resilient P2P FL solutions, and contributes to a deeper understanding of how machine learning interacts with adversarial elements in decentralized networks. \par

In the YAML configuration which is used to drive device setup using Docker, the user can specify if a device is an adversary. The training algorithm (\ref{algo:indevice}) of such a device can be modified to implement attacks like FGSM \cite{goodfellow2014explaining}, RFGSM \cite{tramer2017ensemble}, PGD \cite{madry2017towards} etc. This modelling approach advances the state of the art in P2P FL and helps drive innovation in network security and robustness.\par

If a device is marked as an honest-but-curious node there is no modification to its training algorithm. But a device marked as an adversary can now perform privacy attacks like attribute inference and reconstruction attacks or attacks against robustness like label flipping and evasion attacks.\par

\section{Evaluation}
All the experiments were conducted on Amazon Web Services (AWS) infrastructure using EC2 T2 and M4 instances to host individual devices. The container management and subnet configuration was performed using AWS Elastic Kubernetes Service (EKS) and Virtual Private Cloud (VPC) services respectively. \par
\label{sec:evaluation}
\subsection{Benchmarking Results}
Table \ref{tab:metrics_with_others} presents a comparative analysis of simulation performance with some other simulators widely used in the field. The simulation time for our simulator is comparable to Flower, which has been widely used recently. The learning accuracy achieved is indicated to establish that this is indeed an apples-to-apples comparison, ensuring that we focus on efficiency in computational resource utilization. \par

\begin{table*}[!h]
 \caption{Comparison with other FL simulators}
 \label{tab:metrics_with_others}
    \Centering
        \begin{tabular}{ |c | c | c |c |c |c | }
        \hline
            Simulator & Time(s) & FL Accuracy(\%) \\
            \hline 
            Flower & 812.04 & 69.70  \\
            P2PSim & 911.79 & 68.89  \\
            \textbf{PeerFL} & \textbf{851.15} & 69.61  \\
        \hline
        \end{tabular}
\end{table*}

A notable advantage of PeerFL is its cross-platform compatibility, allowing seamless operation across various operating systems and applying dynamic restrictions during each experiment run. This versatility enables researchers and practitioners to run simulations on various flavours of Ubuntu, Alpine, Raspberry Pi etc., providing flexibility and convenience in different computing environments. \par

\begin{table*}[!h]
 \caption{Performance of PeerFL}
 \label{tab:metrics_with_self}
    \Centering
        \begin{tabular}{ |c | c | c |c |c |c | }
        \hline
            Epochs, Rounds & Number & Model & Time(s) & P2P FL \\
            & of Clients & & & Accuracy(\%) \\
            \hline 
            5, 5 & 2 & 1 Layer NN & 312.19 & 41.33  \\
            5, 5 & 3 & 1 Layer NN & 393.166 & 40.50  \\
            5, 5 & 7 & 1 Layer NN & 460.035 & 39.61  \\
            5, 10 & 10 & VGG-16 & 1260.035 & 69.5  \\
            5, 10 & 10 & VGG-16, Flower & 1201.22 & 69.52  \\
            5, 10 & 10 & Resnet-50 & 1450.41 & 72.63  \\
            5, 5 & 100 & VGG-16 & 82350.1 & 80.41  \\
            5, 5 & 200 & VGG-16 & 10591.83 & 81.20  \\
        \hline
        \end{tabular}
\end{table*}
Table \ref{tab:metrics_with_self} showcases the performance of PeerFL with varying numbers of devices and different neural network model architectures.

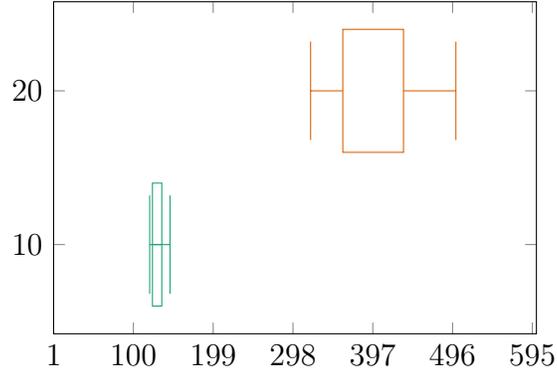
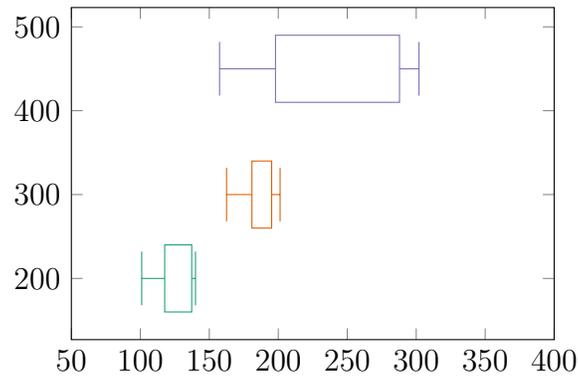
\begin{figure}[H]
\centering
\begin{subfigure}[t]{0.6\textwidth}
\centering
    \begin{tikzpicture}
    \begin{axis} [
        height=6.0cm, width=8.0cm,
        xmin=1, xmax=600, xtick={1,100,...,600},
        ytick={1,2,3},
        yticklabels={10, 20, 30},
    ]
    \addplot+ [
        boxplot prepared={
            upper quartile=123.63,
            lower quartile=135.22,
            upper whisker=145.41,
            lower whisker=120.22,
            draw position=1
        },
    ] coordinates {};
    \addplot+ [
        boxplot prepared={
            upper quartile=435.1,
            lower quartile=359.83,
            upper whisker=500.1,
            lower whisker=319.83,
            draw position=2
        },
    ] coordinates {};
    \end{axis}
\end{tikzpicture}
\caption{Upto 20 simulated devices. X-axis is in seconds.}
\label{fig:boxplot1}
\end{subfigure}

\vspace{\floatsep}

\begin{subfigure}[t]{0.6\textwidth}
    \centering
    \begin{tikzpicture}
    \begin{axis} [
        height=6.0cm, width=8.0cm,
        xmin=6000, xmax=26000,
        xmin=50, xmax=400, xtick={50,100,...,400},
        ytick={1,2,3,4},
        yticklabels={200, 300, 400, 500},
    ]
    \addplot+ [
        boxplot prepared={
            upper quartile=137.25,
            lower quartile=117.66,
            upper whisker=140.00,
            lower whisker=100.99,
            draw position=1
        },
    ] coordinates {};

    \addplot+ [
        boxplot prepared={
            upper quartile=195.2,
            lower quartile=180.91,
            upper whisker=201.33,
            lower whisker=162.51,
            draw position=2
        },
    ] coordinates {};
    
    \addplot+ [
        boxplot prepared={
            upper quartile=287.89,
            lower quartile=197.99,
            upper whisker=302.01,
            lower whisker=157.44,
            draw position=3.5
        },
    ] coordinates {};
    
    \end{axis}
\end{tikzpicture}
\caption{200 to 450 simulated devices. X-axis is in minutes.}
\label{fig:boxplot2}

\end{subfigure}

\caption{Boxplot of variation in simulator run time (in seconds) while scaling up number of devices}
\label{fig:all_boxplot}
\end{figure}

Figure \ref{fig:all_boxplot} shows the amount of time taken to run the simulations with varying numbers of devices. The simulator has been tested to model up to 450 devices simultaneously, as shown in Figure \ref{fig:boxplot2}. For these experiments at scale, the network connectivity graph with bandwidth constraints is generated on the fly. We see an average increase of 47.7 minutes in communication time per every 100 devices added when this underlying graph is sparse and the average out-degree of devices is 3. This drops to 21.3 minutes when the network graph is denser with the average out-degree of devices as 8. Since each device performs its local computations in parallel, the majority of the run time in larger graphs is seen during model transfer time.\par
   
\section{Conclusion and Future Work}
\label{sec:conclusion}
This paper introduces PeerFL, a powerful resource that streamlines the application of peer-to-peer federated learning to address challenges posed by networking. The toolkit seamlessly integrates TensorFlow ML functionality with the ns-3 network, offering a way to leverage network topology modelling to P2P FL problems. The simulator is open-source in nature and encourages community contributions.\par
As part of future work, we plan to extend the portfolio of attack modelling in the simulator so that users can easily conduct experiments that combine different types of adversarial attacks on the simulated network. This in turn will drive our efforts to design defence techniques that can take advantage of the large-scale decentralized nature of peers.\par

\section*{Acknowledgment}
This research is supported by the National Research Foundation, Singapore under its Strategic Capability Research Centres Funding Initiative. Any opinions, findings and conclusions or recommendations expressed in this material are those of the author(s) and do not reflect the views of National Research Foundation, Singapore.



 \bibliographystyle{elsarticle-num} 
 \bibliography{cas-refs}

\begin{thebibliography}{10}
\expandafter\ifx\csname url\endcsname\relax
  \def\url#1{\texttt{#1}}\fi
\expandafter\ifx\csname urlprefix\endcsname\relax\def\urlprefix{URL }\fi
\expandafter\ifx\csname href\endcsname\relax
  \def\href#1#2{#2} \def\path#1{#1}\fi

\bibitem{1902.04885}
Q.~Yang, Y.~Liu, T.~Chen, Y.~Tong, Federated machine learning: Concept and applications (2019).
\newblock \href {http://arxiv.org/abs/arXiv:1902.04885} {\path{arXiv:arXiv:1902.04885}}.

\bibitem{truong2021privacy}
N.~Truong, K.~Sun, S.~Wang, F.~Guitton, Y.~Guo, Privacy preservation in federated learning: An insightful survey from the gdpr perspective, Computers \& Security 110 (2021) 102402.

\bibitem{ASGHARI2019241}
P.~Asghari, A.~M. Rahmani, H.~H.~S. Javadi, \href{https://www.sciencedirect.com/science/article/pii/S1389128618305127}{Internet of things applications: A systematic review}, Computer Networks 148 (2019) 241--261.
\newblock \href {https://doi.org/https://doi.org/10.1016/j.comnet.2018.12.008} {\path{doi:https://doi.org/10.1016/j.comnet.2018.12.008}}.
\newline\urlprefix\url{https://www.sciencedirect.com/science/article/pii/S1389128618305127}

\bibitem{2104.07914}
D.~C. Nguyen, M.~Ding, P.~N. Pathirana, A.~Seneviratne, J.~Li, H.~V. Poor, Federated learning for internet of things: A comprehensive survey (2021).
\newblock \href {http://arxiv.org/abs/arXiv:2104.07914} {\path{arXiv:arXiv:2104.07914}}, \href {https://doi.org/10.1109/COMST.2021.3075439} {\path{doi:10.1109/COMST.2021.3075439}}.

\bibitem{zeng2022federated}
T.~Zeng, O.~Semiari, M.~Chen, W.~Saad, M.~Bennis, Federated learning on the road autonomous controller design for connected and autonomous vehicles, IEEE Transactions on Wireless Communications 21~(12) (2022) 10407--10423.

\bibitem{roy2019braintorrent}
A.~G. Roy, S.~Siddiqui, S.~P{\"o}lsterl, N.~Navab, C.~Wachinger, Braintorrent: A peer-to-peer environment for decentralized federated learning, arXiv preprint arXiv:1905.06731 (2019).

\bibitem{kairouz2021advances}
P.~Kairouz, H.~B. McMahan, B.~Avent, A.~Bellet, M.~Bennis, A.~N. Bhagoji, K.~Bonawitz, Z.~Charles, G.~Cormode, R.~Cummings, et~al., Advances and open problems in federated learning, Foundations and Trends{\textregistered} in Machine Learning 14~(1--2) (2021) 1--210.

\bibitem{doi:10.1080/13600834.2019.1573501}
C.~J. Hoofnagle, B.~van~der Sloot, F.~Z. Borgesius, \href{https://doi.org/10.1080/13600834.2019.1573501}{The european union general data protection regulation: what it is and what it means}, Information \& Communications Technology Law 28~(1) (2019) 65--98.
\newblock \href {http://arxiv.org/abs/https://doi.org/10.1080/13600834.2019.1573501} {\path{arXiv:https://doi.org/10.1080/13600834.2019.1573501}}, \href {https://doi.org/10.1080/13600834.2019.1573501} {\path{doi:10.1080/13600834.2019.1573501}}.
\newline\urlprefix\url{https://doi.org/10.1080/13600834.2019.1573501}

\bibitem{CHIK2013554}
W.~B. Chik, \href{https://www.sciencedirect.com/science/article/pii/S0267364913001374}{The singapore personal data protection act and an assessment of future trends in data privacy reform}, Computer Law \& Security Review 29~(5) (2013) 554--575.
\newblock \href {https://doi.org/https://doi.org/10.1016/j.clsr.2013.07.010} {\path{doi:https://doi.org/10.1016/j.clsr.2013.07.010}}.
\newline\urlprefix\url{https://www.sciencedirect.com/science/article/pii/S0267364913001374}

\bibitem{privacy_gaff}
B.~Gaff, H.~Sussman, J.~Geetter, Privacy and big data, Computer 47 (2014) 7--9.
\newblock \href {https://doi.org/10.1109/MC.2014.161} {\path{doi:10.1109/MC.2014.161}}.

\bibitem{mcmahan2017communicationefficient}
B.~McMahan, E.~Moore, D.~Ramage, S.~Hampson, B.~A.~y. Arcas, \href{https://proceedings.mlr.press/v54/mcmahan17a.html}{{Communication-Efficient Learning of Deep Networks from Decentralized Data}}, in: A.~Singh, J.~Zhu (Eds.), Proceedings of the 20th International Conference on Artificial Intelligence and Statistics, Vol.~54 of Proceedings of Machine Learning Research, PMLR, 2017, pp. 1273--1282.
\newline\urlprefix\url{https://proceedings.mlr.press/v54/mcmahan17a.html}

\bibitem{2203.12285}
Z.~Li, J.~Lu, S.~Luo, D.~Zhu, Y.~Shao, Y.~Li, Z.~Zhang, Y.~Wang, C.~Wu, Towards effective clustered federated learning: A peer-to-peer framework with adaptive neighbor matching (2022).
\newblock \href {http://arxiv.org/abs/arXiv:2203.12285} {\path{arXiv:arXiv:2203.12285}}, \href {https://doi.org/10.1109/TBDATA.2022.3222971} {\path{doi:10.1109/TBDATA.2022.3222971}}.

\bibitem{2101.12316}
N.~Gupta, N.~H. Vaidya, Byzantine fault-tolerance in peer-to-peer distributed gradient-descent (2021).
\newblock \href {http://arxiv.org/abs/arXiv:2101.12316} {\path{arXiv:arXiv:2101.12316}}.

\bibitem{2007.14390}
D.~J. Beutel, T.~Topal, A.~Mathur, X.~Qiu, J.~Fernandez-Marques, Y.~Gao, L.~Sani, K.~H. Li, T.~Parcollet, P.~P.~B. de~Gusmão, N.~D. Lane, Flower: A friendly federated learning research framework (2020).
\newblock \href {http://arxiv.org/abs/arXiv:2007.14390} {\path{arXiv:arXiv:2007.14390}}.

\bibitem{2104.03042}
A.~Mathur, D.~J. Beutel, P.~P.~B. de~Gusmão, J.~Fernandez-Marques, T.~Topal, X.~Qiu, T.~Parcollet, Y.~Gao, N.~D. Lane, On-device federated learning with flower (2021).
\newblock \href {http://arxiv.org/abs/arXiv:2104.03042} {\path{arXiv:arXiv:2104.03042}}.

\bibitem{liu2021fate}
Y.~Liu, T.~Fan, T.~Chen, Q.~Xu, Q.~Yang, Fate: An industrial grade platform for collaborative learning with data protection, The Journal of Machine Learning Research 22~(1) (2021) 10320--10325.

\bibitem{githubGitHubFacebookresearchFLSim}
{G}it{H}ub - facebookresearch/{F}{L}{S}im: {F}ederated {L}earning {S}imulator ({F}{L}{S}im) is a flexible, standalone core library that simulates {F}{L} settings with a minimal, easy-to-use {A}{P}{I}. {F}{L}{S}im is domain-agnostic and accommodates many use cases such as vision and text. --- github.com, \url{https://github.com/facebookresearch/FLSim}, [Accessed 06-01-2024].

\bibitem{article}
L.~Campanile, M.~Gribaudo, M.~Iacono, F.~Marulli, M.~Mastroianni, Computer network simulation with ns-3: A systematic literature review, Electronics 9 (2020) 272.
\newblock \href {https://doi.org/10.3390/electronics9020272} {\path{doi:10.3390/electronics9020272}}.

\bibitem{inbook}
P.~Jako, Boxing, 2009, pp. 1--21.
\newblock \href {https://doi.org/10.1007/978-1-84800-354-5_12} {\path{doi:10.1007/978-1-84800-354-5_12}}.

\bibitem{prechelt2002early}
L.~Prechelt, Early stopping-but when?, in: Neural Networks: Tricks of the trade, Springer, 2002, pp. 55--69.

\bibitem{goodfellow2014explaining}
I.~J. Goodfellow, J.~Shlens, C.~Szegedy, Explaining and harnessing adversarial examples, arXiv preprint arXiv:1412.6572 (2014).

\bibitem{tramer2017ensemble}
F.~Tram{\`e}r, A.~Kurakin, N.~Papernot, I.~Goodfellow, D.~Boneh, P.~McDaniel, Ensemble adversarial training: Attacks and defenses, arXiv preprint arXiv:1705.07204 (2017).

\bibitem{madry2017towards}
A.~Madry, A.~Makelov, L.~Schmidt, D.~Tsipras, A.~Vladu, Towards deep learning models resistant to adversarial attacks, arXiv preprint arXiv:1706.06083 (2017).

\end{thebibliography}





\end{document}